\documentclass[a4paper,fleqn,usenatbib,useAMS]{mnras}

\usepackage{graphicx}	
\usepackage{amsmath}	
\usepackage{amssymb}	
\usepackage{multicol}        
\usepackage{bm}		
\usepackage{pdflscape}	
\usepackage{color}

\usepackage[T1]{fontenc}
\usepackage{ae,aecompl}

\usepackage{newtxtext,newtxmath}

\title[Cloud and Aurora from KLCAM]{Cloud Cover and Aurora Contamination 
at Dome~A in 2017 from KLCAM}

\author[Xu Yang. et al.]{Xu Yang$^{1,2}$, Zhaohui Shang$^{1,3}$\thanks{Contact e-mail: \href{mailto:zshang@gmail.com}{:zshang@gmail.com}},Keliang Hu$^{1}$, Yi Hu$^{1}$, Bin Ma$^{1}$, Yongjiang Wang$^{1,2}$, 
\newauthor Zihuang Cao$^1$, Michael C.B. Ashley$^{4}$ , Wei Wang$^{1}$
\\
\\$^{1}$National Astronomical Observatories, Chinese Academy of Sciences, Beijing 100101, China
\\$^{2}$University of Chinese Academy of Sciences, Beijing 100049, China
\\$^{3}$Tianjin Normal University, Tianjin 300387, China
\\$^{4}$School of Physics, University of New South Wales, Sydney NSW 2052, Australia}

\date{Accepted XXX. Received YYY; in original form ZZZ}

\pubyear{2020}

\begin{document}
\label{firstpage}
\pagerange{\pageref{firstpage}--\pageref{lastpage}}
\maketitle

\begin{abstract}
Dome A in Antarctica has many characteristics that make it an excellent site for astronomical
observations, from the optical to the terahertz.  Quantitative site testing is still needed to confirm
the site's properties.  
In this paper, we present a statistical analysis of cloud cover and aurora 
contamination from the Kunlun Cloud and Aurora Monitor (KLCAM).
KLCAM is an automatic, unattended all-sky camera aiming for long-term
monitoring of the usable observing time and optical sky background at Dome~A.  
It was installed at Dome~A in January 2017, worked through
the austral winter, and collected over 47,000 images over 490 days.
A semi-quantitative visual data analysis of cloud cover and auroral
contamination was carried out by five individuals.  
The analysis shows that the night sky was free of clouds
for 83 per cent of the time, which ranks Dome~A highly in a comparison with other observatory sites.
Although aurorae were detected somewhere on an image for nearly 45 per cent of the time, the chance of a point on the sky
being affected by an aurora is small. The
strongest auroral emission lines can be filtered out with customized
filters.
\end{abstract}




\section{Introduction}

Dome A is an excellent ground-based astronomical site at the highest point
on the Antarctic plateau at an altitude of 4,093\,m above sea level.  It was first visited by the
Chinese National Antarctic Research Expedition (CHINARE) in 2005 and
is also the place where the Chinese Kunlun Station is located. 
Various telescopes and astronomical instruments have been
installed at Dome~A over a 10 year period for both astronomical observations and site
testing \citep{Shang20}.

Given the harsh environment at Dome~A and the fact that Kunlun Station
is not currently a winterover station, the development and operation of
the telescopes and instruments at Dome~A is a great challenge.  The equipment must
be designed to work unattended and fully automatically for at least a
year, with the capability of remote control via very limited bandwidth of Iridium satellite.  
A good example is the Antarctic Survey Telescopes
for time-domain astronomy \citep[AST3,][]{Ma20a}; the instruments for
site testing face the same challenges and requirements.

Many site testing instruments have been installed and operated at
Dome~A, and the results have shown that Dome~A is an
exceptional site for astronomical observations.  
The Surface layer Non-Doppler
Acoustic Radar (SNODAR; \citealt{snodar-ins}) showed that the median thickness of
the atmospheric turbulent boundary layer is remarkably thin at 13.9\.m \citep{snodar}, allowing the excellent seeing in
the free atmosphere to be accessed from a tower of modest height.
\citet{Ma20b} recently confirmed this using the combined data from the
KunLun Differential Motion Monitor (KL-DIMM; \citealt{kl-dimm}) on an
8\,m tower and the multi-layer Kunlun Automated Weather
Station \citep{Hu14,klaws} on a 15\,m tower.  KL-DIMM observed
seeing as low as 0.13\arcsec, with a median of
0.31\arcsec\ \citep{Ma20b} when the boundary layer was below the telescope's aperture.
In the terahertz (sub-millimetre) regime, instruments such as the
prototype High Elevation Antarctic Terahertz telescope (Pre-HEAT;
\citealt{pre-heat}), Nigel \citep{nigel-water}, and the Fourier
Transform Spectrometer (FTS; \citealt{fts}) have shown the perceptible
water vapour at Dome~A is extremely low, less than half a millimetre, making the terahertz sky transparency exceptionally good. 
Other instruments such as Gattini \citep{gattini}, Nigel \citep{nigel},
and the Chinese Small Telescope ARray (CSTAR; \citealt{zouhu}) have
studied the sky background, airglow, and aurora at Dome~A.
A recent detailed review of astronomy at Dome~A, including site
testing, can be found in
\citet{Shang20}.

Among the key parameters of astronomical site testing, the fraction of
clear nights free of clouds is crucial for optical and infrared
astronomy. This is particularly important at Dome~A where the Antarctic winter
provides the opportunity to obtain continuous observations spanning days and weeks.  
Using 5 months of CSTAR data from Dome~A during 2008 winter to define a
relative transparency variation, \citet{zouhu} reported a photometric
night (extinction < 0.3 mag) fraction of 67 per cent. The
field-of-view of CSTAR was 20\,deg$^2$, centred on the south
celestial pole which is about 10 degrees from local zenith.   
\citet{gattini-cloud} also estimated the cloud cover at Dome~A with
Gattini data of 2009.  They measured the sky transparency of images
and obtained a photometric night fraction of 62.4 per cent, roughly
consistent with the CSTAR results, but in a much wide field of view of
$90^{\circ}$ × $90^{\circ}$, which was also centred on the south
celestial pole.  
These results show that the photometric night fraction at Dome~A is
very promising compared to Mauna Kea or northern Chile \citep{zouhu}.
However, the data covered only part of the sky and long-term, systematic
monitoring was not available.

Several different methods have been developed over the past decades to carry
out systematic estimates of clear night fraction.  In the early years
of site testing, such as for the Very Large Telescope Project (VLT;
\citealt{VLT}), a visual estimate of cloud cover was used.  
As technology evolved, satellite data reference have also been used to analyse
cloud cover in site testing for large projects such as the European
Extreme Large Telescope (E-ELT; \citealt{ELT}) and the Thirty Meter
Telescope (TMT; \citealt{TMT}).  However, the results from satellite
data were proven to be reliable only after compared with and verified
by ground-based observations \citep{TMT}. Furthermore, satellite images over Antarctica during wintertime have difficulty
distinguishing the difference between cloud and the ice surface, since both are a similar temperature and colour.
Nowadays, all-sky cameras are usually the best option for ground-based observations of
cloud cover, with the obvious advantages that images are recorded frequently
and can be analysed systematically. In the case of the site testing for TMT, the All-Sky CAmera (ASCA)
was used for measuring cloud cover and calibrating the satellite data
\citep{asca}.  

At Dome~A, an earlier all-sky camera, the High-Resolution CAMera
(HRCAM) \citep{hrcam} obtained data throughout 2010 \citep{simsthesis}, but coverage in later years was affected
by shutter and disk drive failures.
To improve the coverage we designed the Kunlun Cloud and Aurora Monitor (KLCAM) to
continue the study of both cloud cover and aurora contamination at Dome~A.

In Section~\ref{sec:ins} of this article we present the basic design of KLCAM and the
pre-deployment tests we undertook to ensure that KLCAM would work at Dome~A.
In Section~\ref{sec:data}, we describe the
observation strategy we used with KLCAM, and the data we obtained.
The method we use to analyse the data and the results 
are described in Section~\ref{sec:res}.  In Section~\ref{sec:dis}, we
discuss some issues of the work, the advantages and disadvantages of
our analysis, and future work.

\section{Instrument}
\label{sec:ins}

The detailed design of KLCAM can be found in \citet{klcam1}.  Here
we briefly present the basic features of KLCAM.

\begin{figure}
\includegraphics[width=\columnwidth]{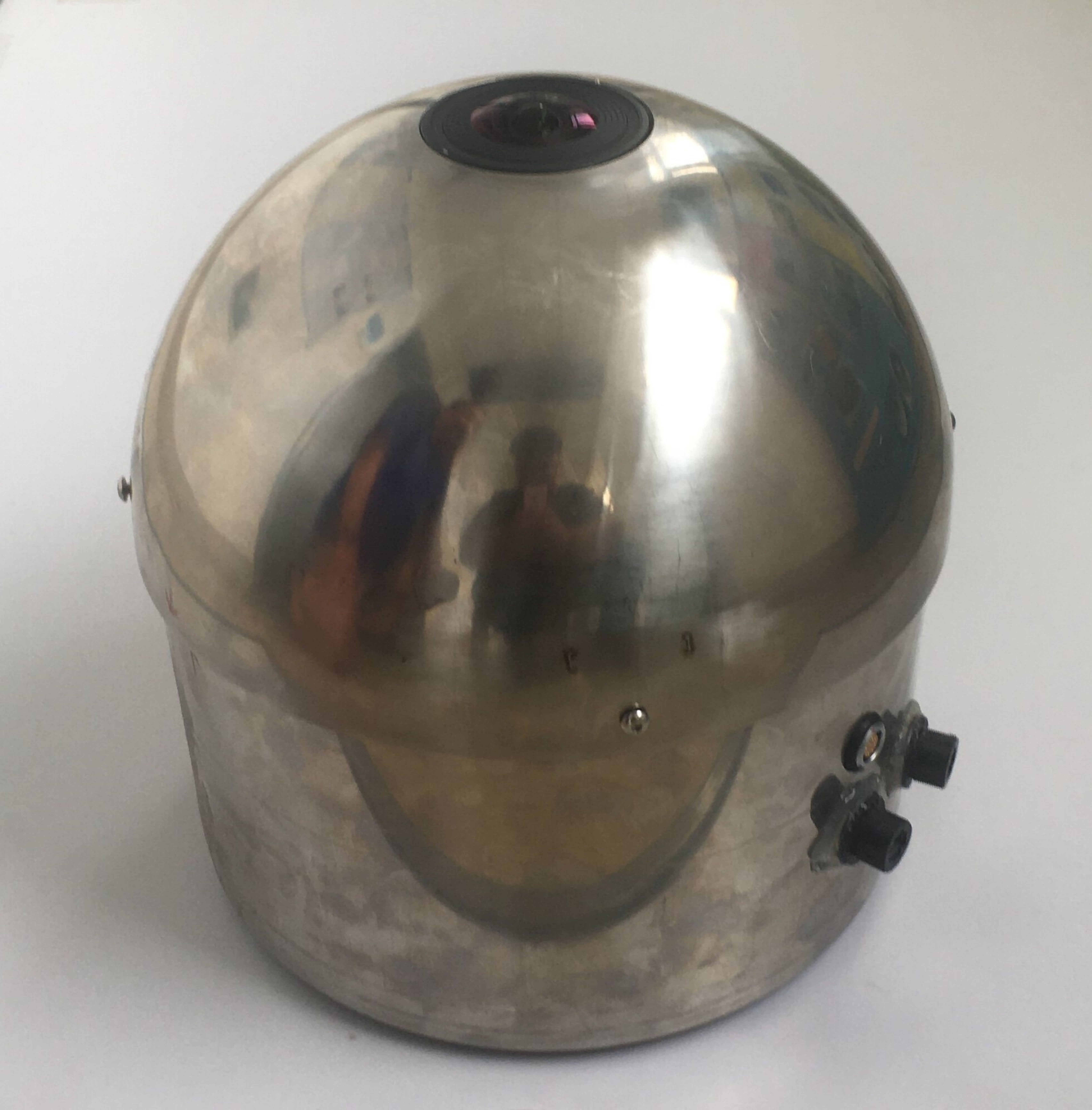}
\caption{A photo of KLCAM.  On the top is the camera lens.
The sockets on the side of the metal shell 
are for the network connection and power supply.}
\label{fig:KLCAM1}
\end{figure}

KLCAM has a Canon 100D camera equipped with a Sigma 4.5mm f/2.8 fish-eye
lens that allows complete coverage of the sky from the zenith to the horizon in all directions.
KLCAM is controlled by a customized ARM-based computer.
Fig.~\ref{fig:KLCAM1} shows the body of KLCAM with everything inside
the metal shell. The thermal design of KLCAM allowed it to work at the low temperatures and low atmospheric pressures at Dome~A.  
The camera sits on an isothermal plate, with a thick insulation layer of low thermal
conductivity material between the camera and the low emissivity metal shell. Similarly to the earlier HRCAM, the thermal
design forces most of the internal heat to pass through the fish-eye lens, thereby
preventing frosting on the lens. 

An active heating system keeps the camera temperature
between $0\degr$C and $10\degr$C.  Because of the good insulation,
only 10 watts is needed for heating, even when
the ambient temperature drops as low as $-80\degr$C.  In the case of
unexpected problems, we can access the camera remotely via satellite
communication.

Before being deployed to Dome~A, KLCAM was tested both in the laboratory and in the
field.
We tested KLCAM down to $-70\degr$C inside a cold chamber
and verified that the thermal design could keep the camera temperature
above $0\degr$C.
To verify the performance of KLCAM in the low air pressure and
unattended environment of Dome~A, we tested it at a site with an
altitude of 4,500 metres above sea level.  KLCAM worked perfectly
during all of these tests and  was installed at Dome~A in 2017 by the
33rd CHINARE team.  The power and Iridium satellite communication for
KLCAM as well as other instruments were supported by the automated observatory platform PLATO-A
\citep{plato}.

\section{Observations and Data}
\label{sec:data}

\begin{figure}
 \includegraphics[width=\columnwidth]{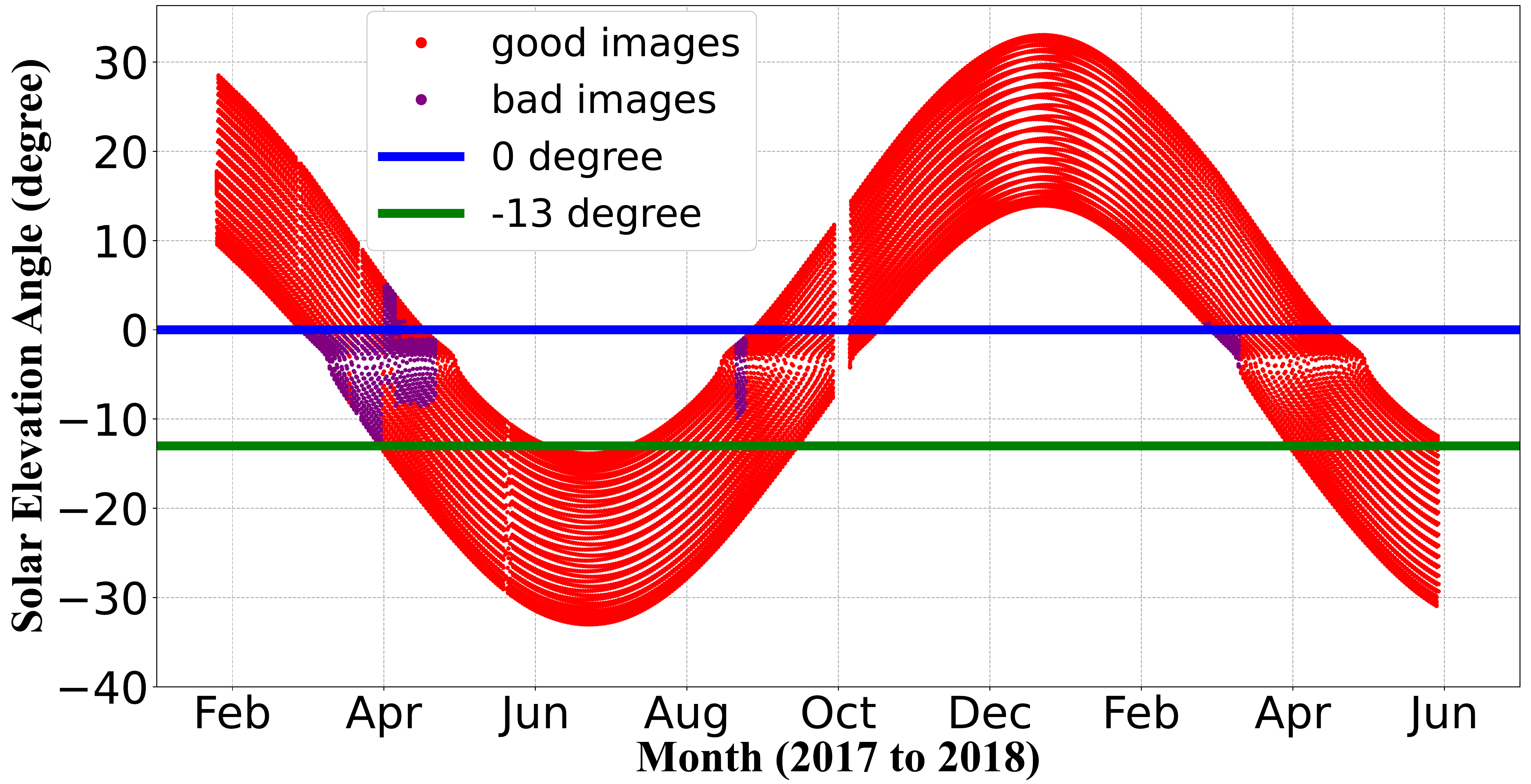}
 \caption{The KLCAM data points shown in the plot of solar elevation angle
vs. observation date. Each red dot represents one good image, and
each purple dot represents one bad image. The blue line
indicates the horizon, and the green line marks the solar 
elevation angle of $-$13\degr\,. 
There was a 5-day downtime, shown as the small gap in
October.
}
 \label{fig:datapoints}
\end{figure}

From the date of installation at Dome~A in January 2017, KLCAM worked well
for 490 days except for a 5-day downtime in October 2017 due
to a power issue, shown as the small gap in Fig.~\ref{fig:datapoints}.
KLCAM worked unattended throughout the whole Antarctic winter until
PLATO-A ran out of fuel in May 2018. 

We developed an observing strategy to balance the frequency of the
monitoring and the camera shutter life.  The camera of KLCAM is a
commercial digital camera with a mechanical shutter which is a hidden
danger for the system in the harsh environment.  In addition, KLCAM
has to work in a real unattended situation at Dome~A, without
repairing or replacement for at least one year, and sometimes 2 years if
there is no Dome~A expedition by CHINARE for that year.
Considering these factors, we decided to take an exposure every 30
minutes to ensure a two-year lifetime.  In order to obtain proper exposures, 
the exposure time was time dependant, which is determined by the
elevation angles of the Sun and the Moon, along with the phase of the Moon.
We also changed the ISO, the sensitivity of the camera sensor, during the daytime or full moon to prevent
overexposure.  During dawn and dusk, since the sky brightness changes
rapidly, we took multiple images with different exposure times
as an extra safety to avoid over- or underexposure.

The images were saved in a Solid State
Drive (SSD) inside KLCAM itself, and then downloaded to our storage
system installed inside PLATO-A as backups \citep{Ma20a}.  
From 2017 Jan 17 to 2018 May 28 before PLATO-A ran out of fuel,
a total of 47,035 images were
taken covering 485 days (excluding a 5-day gap) as shown in
Fig.~\ref{fig:datapoints}.  However, here we are only interested in
nighttime images, defined as those taken between when twilight started
and when twilight ended.  At Dome~A, twilight occurs when the sun is
13\degr\, below the horizon (see \citealt{zouhu}).
We first excluded 1,062 saturated or underexposed bad images,
including only two nighttime images at the very beginning of the
season.  
When there were multiple exposures at a time, we only kept the best
one, excluding 23,953 images in total.  After these selections, there
are 22,020 images left, of which only 6,664 nighttime images are used
in the following analysis.

\section{Analysis and results}
\label{sec:res}

To analyse the cloud cover and aurora contamination at Dome~A, we
adopted a method similar to that for TMT site testing to classify the
all-sky images by visual analysis.

\subsection{TMT Method}

\begin{figure}
 \includegraphics[width=\columnwidth]{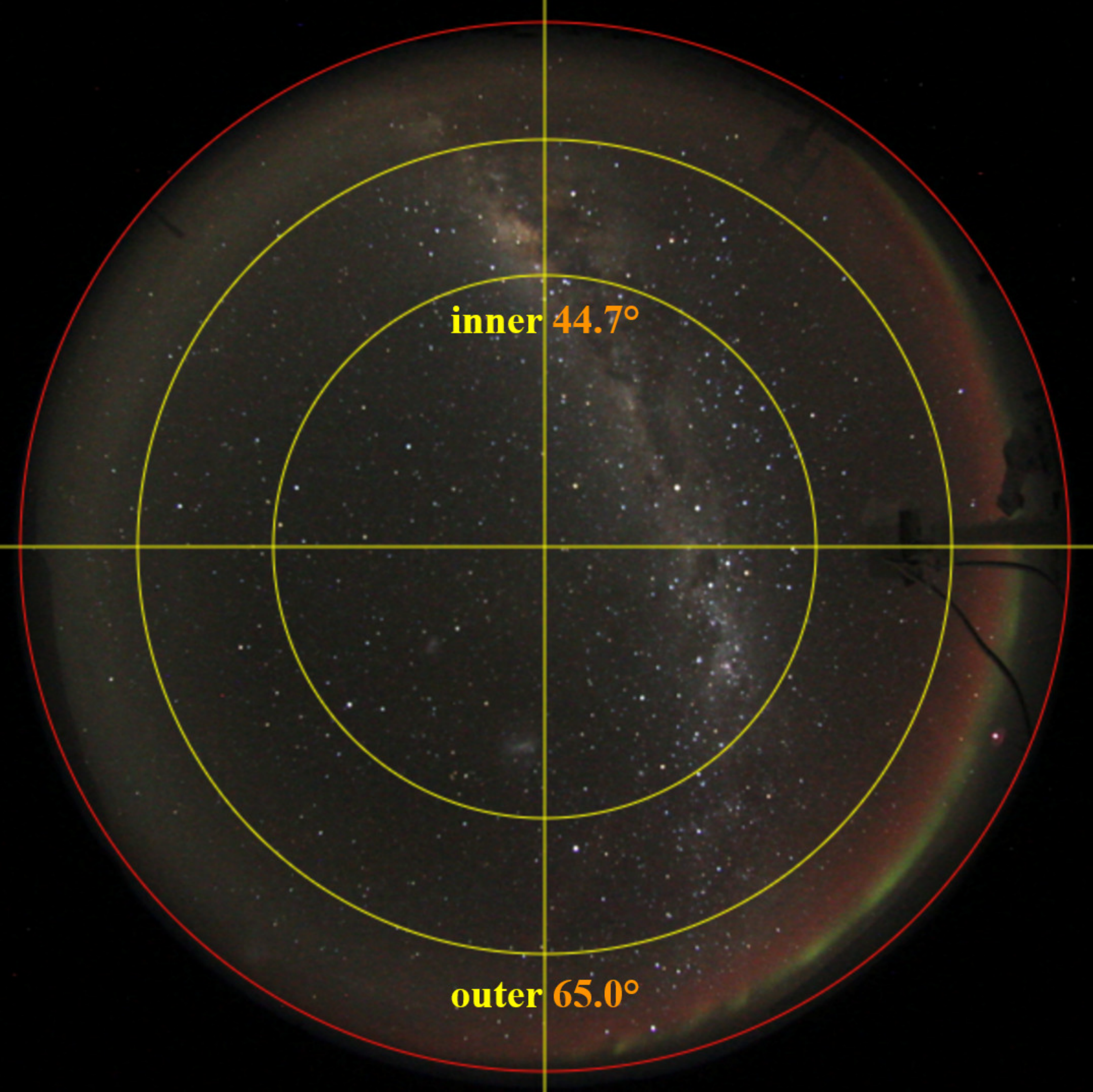}
 \caption{
Regions defined on a KLCAM image based on the visual inspection
method for TMT site testing.  The yellow circles indicate zenith
angles of 44.7\degr\ and 65\degr, respectively.  
The red circle marks the horizon.
}
 \label{fig:sample}
\end{figure}

Site testing for TMT used a visual analysis method of all-sky image, due to the complexity of
analysing the images quantitatively.
First, they divided each all-sky image into three regions used two
circles centred on the local zenith, with the outer one at a zenith angle of
65\degr\ and the inner one at 44.7\degr\ (Fig.~\ref{fig:sample}).  The angles were chosen so that
the outer region, which is the annulus between the two
circles, has the same sky area as the inner circle, and the third region
outside the outer circle is simply ignored because of the observing limit
of TMT.

The images were then made into movies, each of which covering an hour
of nighttime data. Independent analysers then visually inspected and
classified the movies into one of the following four pre-defined classes
\citep{asca}:

\begin{itemize}

\item \verb'clear' -- No clouds are detected inside the 65\degr\
zenith angle circle.

\item \verb'outer' -- Clouds are only detected inside the outer ring
(between 65\degr\ and 44.7\degr\ circles).

\item \verb'inner' -- Clouds are detected inside the inner 44.7\degr\
circle.

\item \verb'cover (cloudy)' -- Over half of the area inside the 65
\degr\ circle are covered with clouds.

\end{itemize}

Finally, the clear (cloud-free) fraction of a site were derived from the
statistics on the classification of all data.

\subsection{Our Modified Method}
\label{sec:ourmethod}

Similar to the TMT method, we have defined four categories of cloud
cover for analysing KLCAM images.  Instead of using movies, we use
individual images, each of which represents half an hour.

As the categories are ordered in a sequence of increasing cloud cover,
we assign a number of 1 to 4 to each category, respectively, and
therefore can do the analysis and error estimate semi-quantitatively.

Each image was inspected, classified, and marked independently by five
individuals, and the median value was taken as the resultant category
of the image.
The statistical error can be given by the confusion matrix
(Section~\ref{sec:aurora}).  The matrix shows the discrepancy in the
classification results from different analysers.  
For example, if one image is marked as `clear' by four analysers, and
as `outer' by one analyser, its median value will put it into the
result class of `clear'.  In the confusion matrix, there are 4 counts
in the marked class of `clear' and 1 count in `outer', indicating a 20
per cent disagreement with the result class of `clear'. 

Another difference from TMT site testing is that we also need to
consider aurora.  However, since the cloud and aurora are independent
of each other, we can analyse them separately in a similar way, 
using four categories for aurora contamination.

\subsection{Aurora}
\label{sec:aurora}

\begin{figure}
 \includegraphics[width=\columnwidth]{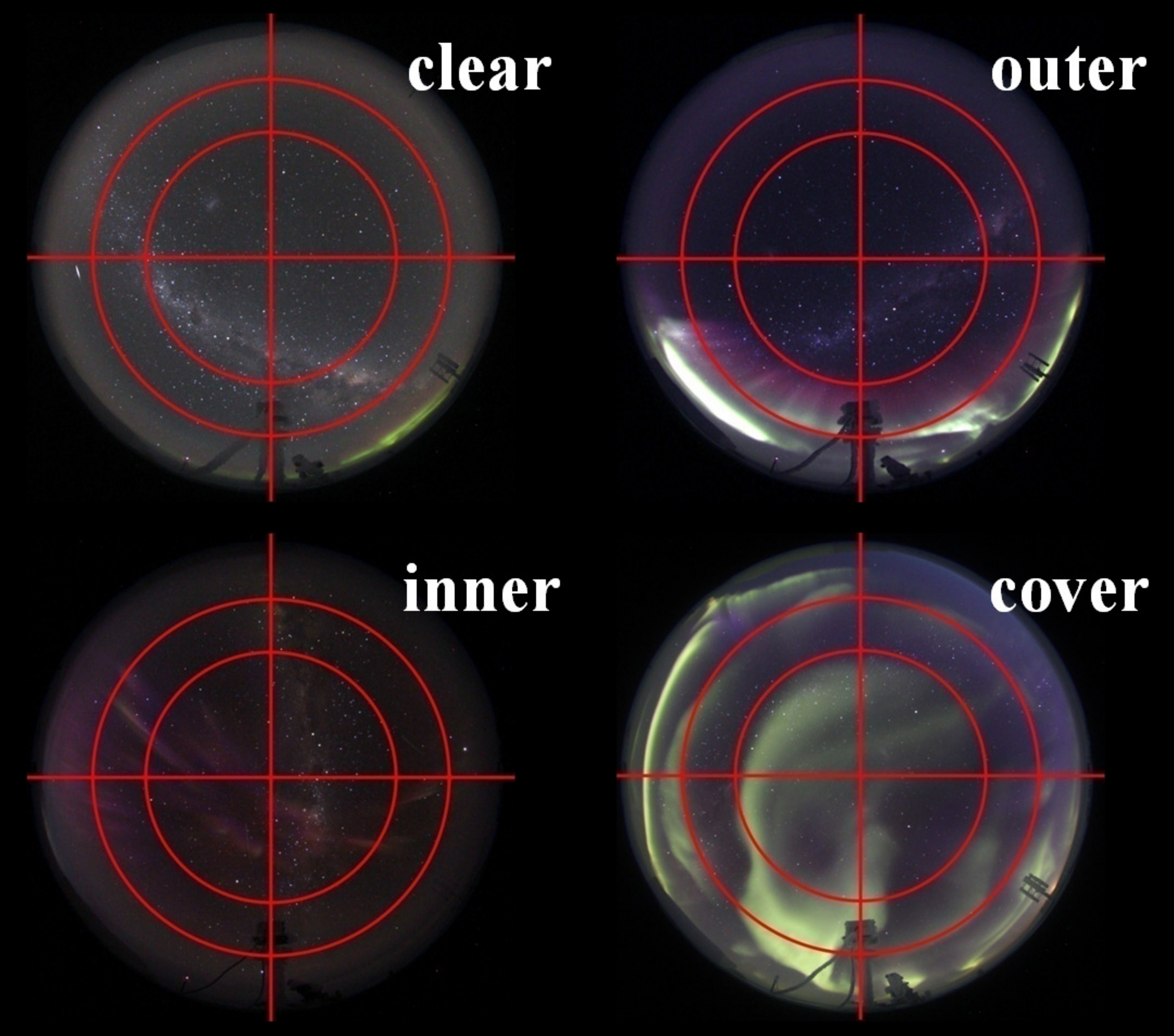}
 \caption{Examples of aurora classification.
}
 \label{fig:auroraclass(small)}
\end{figure}

We analyse aurorae using a similar classification
as that of TMT site testing for clouds. 
Fig.~\ref{fig:auroraclass(small)} shows examples of the aurora
classification described below.

\begin{itemize}

\item \verb'clear' -- No aurorae are detected inside the 65\degr\
zenith angle circle.

\item \verb'outer' -- Aurorae are only detected inside the outer ring
(between 65\degr\ and 44.7\degr\ circle).

\item \verb'inner' -- Aurorae are detected inside the inner
44.7\degr\ circle.

\item \verb'cover' -- Over half of the area inside the 65\degr\ circle
are covered with aurorae.

\end{itemize}

\begin{figure}
 \includegraphics[width=\columnwidth]{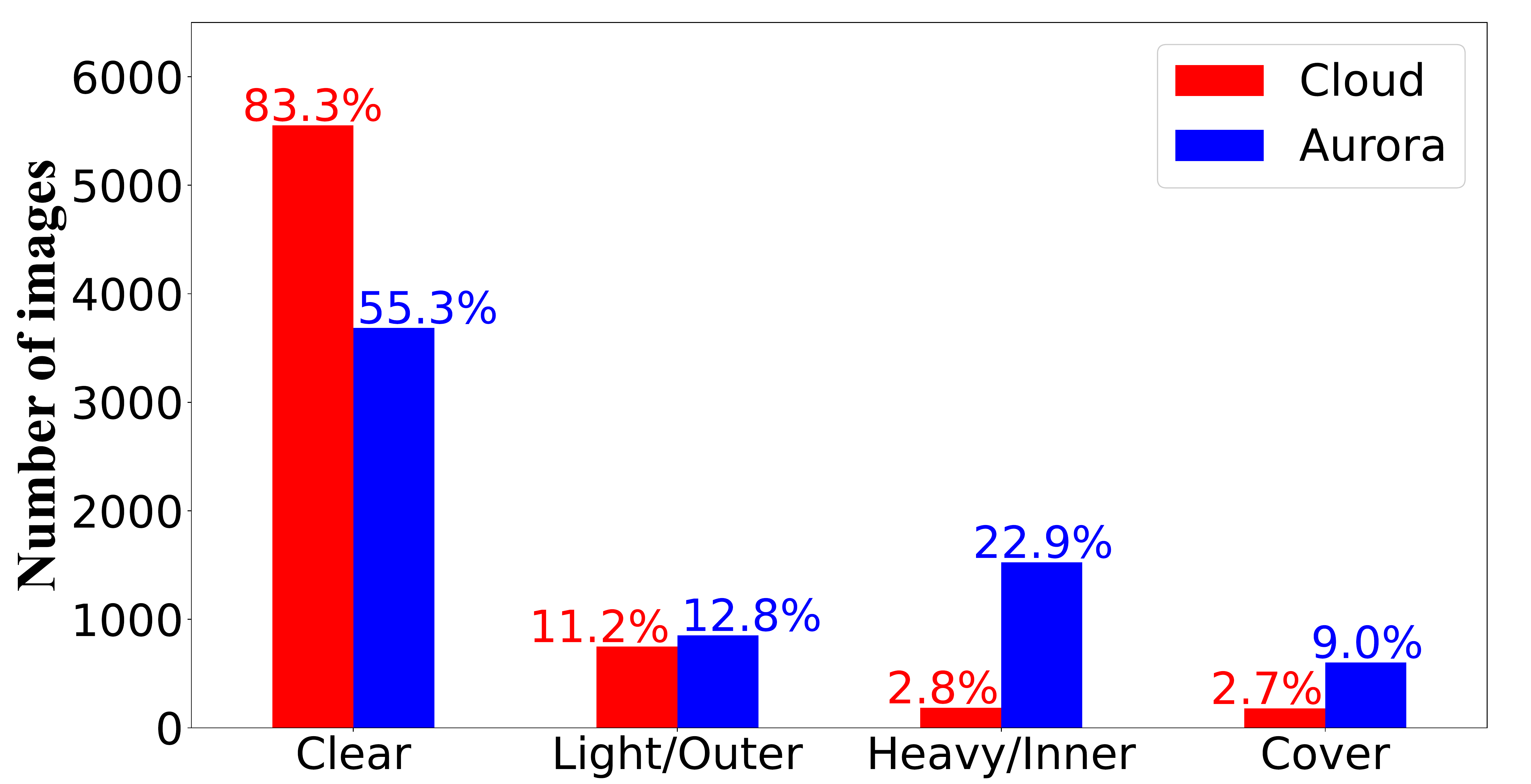}
 \caption{Statistical results from visual analysis of cloud and aurora.}
 \label{fig:result}
\end{figure}

We present the statistical results for aurora classification in
Fig.~\ref{fig:result} together with the results for cloud cover
(Section~\ref{cloud}).
Our results shows that the sky is free of aurorae for about 55 per cent of the
time, as indicated by the `clear' class.  Strong aurorae, marked as
`cover' class, happened for 9 per cent of the time, indicating that
such aurorae are not common at Dome~A.  The remaining 36 per
cent of `inner' and `outer' classes were usually relatively
faint aurorae that are common, as can be seen in
Fig.~\ref{fig:auroraclass(small)}.  This agrees with the spectroscopic
data from NIGEL \citep{nigel}.  Since aurorae are composed of
only emission lines, customized filters can easily exclude the strong auroral
emission lines so as to minimize their contamination to the sky
background. Since aurorae only occupy a fraction of the sky, the probability of 
an observation made with a narrow-field telescope being
affected is small.
For example, even with an FOV of $4.5\degr$~$\times$~$4.5\degr$,
CSTAR images, centered on the south celestial pole, 
were only affected by aurorae for 2 per cent of 
the time during the 4-month observing season in 2008 \citep{zouhu}. 
However, we did not attempt to estimate this 
probability in this semi-quantitative work and our
future work will study the spatial distribution and probability
of the aurorae (Section~\ref{sec:dis}).

\begin{figure}
 \includegraphics[width=\columnwidth]{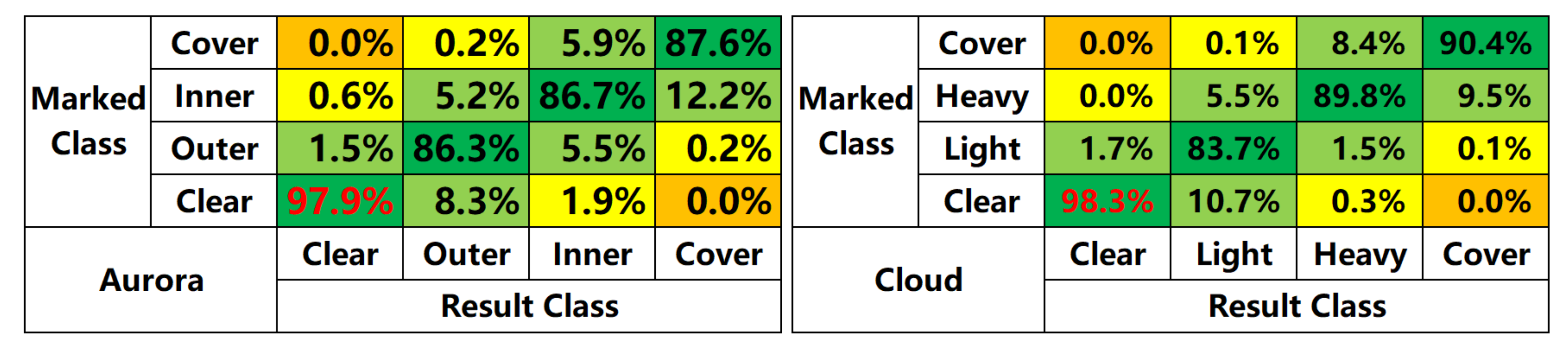}
 \caption{
Confusion matrices of the statistics for aurora (left) and cloud
(right), showing the uncertainties in the result classes due to differences
in the classifications from our five human analysers.
}
 \label{fig:matrix(small)}
\end{figure}

Fig.~\ref{fig:matrix(small)} shows the confusion matrix for our
uncertainty estimates for aurorae.  When there was no aurora, the classifications
from our five analysers agree at a level of 98 per cent, indicating
a very reliable result.  However, when aurorae occurred in the result
classes of `outer', `inner', and `cover', the agreements dropped to
as low as 86 per cent.  For example, in the statistical results for images
classified as the `cover' class, about 12 per cent of them were also marked as
the `inner' class by some analysers.
These disagreements result from the fact that some faint aurorae are
hard to detect.

As expected, the agreement between the marked
classes and a certain result class decreases when the marked one
is further from the result one,
since each result class actually represents the median of
the marked classes as described in Section~\ref{sec:ourmethod}.  This
further demonstrates the robustness of our method.

\subsection{Cloud}
\label{cloud}

The situation for cloud analysis is a little different as we cannot
use the TMT method since the clouds at Dome~A are qualitatively different.  Most
of the clouds at Dome~A are diffuse, so it is hard to classify them as `inner'
or `outer'.  Moreover, one cannot always classify them as `cover' since
there is virtually no real overcast sky and the clouds
simply just increase extinction.
This was also reported from CSTAR data, where `cloudy' or worse
situation accounted for only 2 per cent of the time, compared to 30
per cent at Mauna Kea \citep{zouhu}.
Therefore, we defined a different classification scheme below, in an attempt
to semi-quantitatively characterize the extinction:

\begin{itemize}

\item \verb'clear' -- No clouds are detected inside the 65\degr\
zenith angle circle.

\item \verb'light' -- Marginal or minor extinction can be inferred, but the Milky Way
is still clear and the stars are bright. 

\item \verb'heavy' -- High extinctions can be inferred, the number of
stars decreases dramatically, or the shape of clouds can be clearly
seen.

\item \verb'cover' -- Clouds cover all the area and few or no stars can be
seen.

\end{itemize}

\begin{figure}
 \includegraphics[width=\columnwidth]{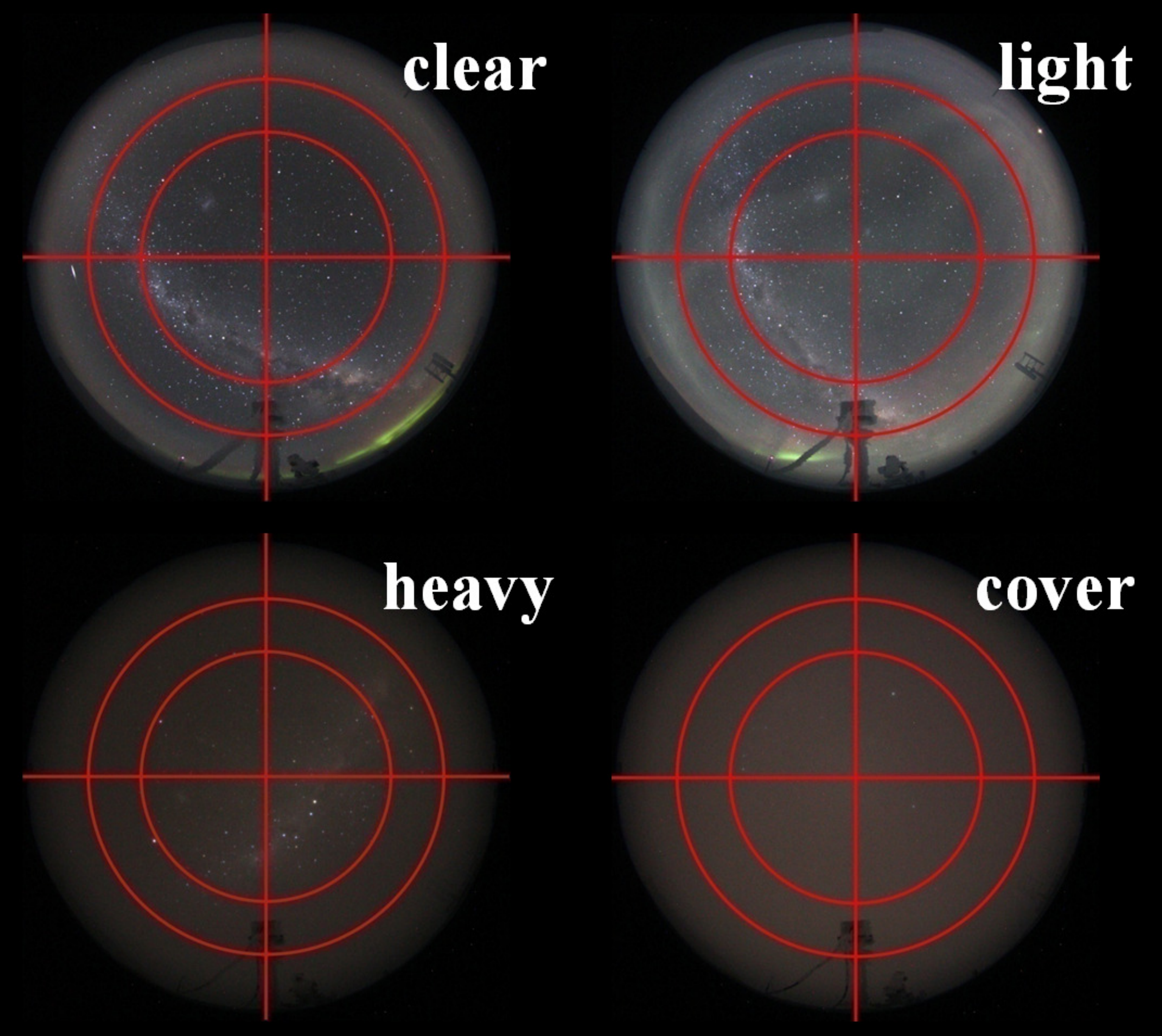}
 \caption{Examples of cloud classification. 
}
 \label{fig:cloudclass(small)}
\end{figure}

Fig.~\ref{fig:cloudclass(small)} shows examples of our cloud
classification and Fig.~\ref{fig:result} shows the statistical
results.  The night sky at Dome~A was free of clouds over 83 per cent
of the time, and the worst case with `cover' clouds was only about 3 per
cent, consistent with CSTAR results \citep{zouhu}.

The confusion matrix for cloud statistics in Fig.~\ref{fig:result}
shows that the classification for `clear' class is very reliable, with
an uncertainty of only 2 per cent among the analysers.  
The `heavy' and `cover' classes have  better agreement than
corresponding classes in aurora classification, because the effect of
heavy extinction on images is easier to be detected by eye than
faint aurorae.  However, this is not the case for the `light' class in
which different analysers seem to be more inconsistent (about 16 per
cent).
We attribute this as a subjective effect as the method is semi-quantitative after all.

\begin{figure}
 \includegraphics[width=\columnwidth]{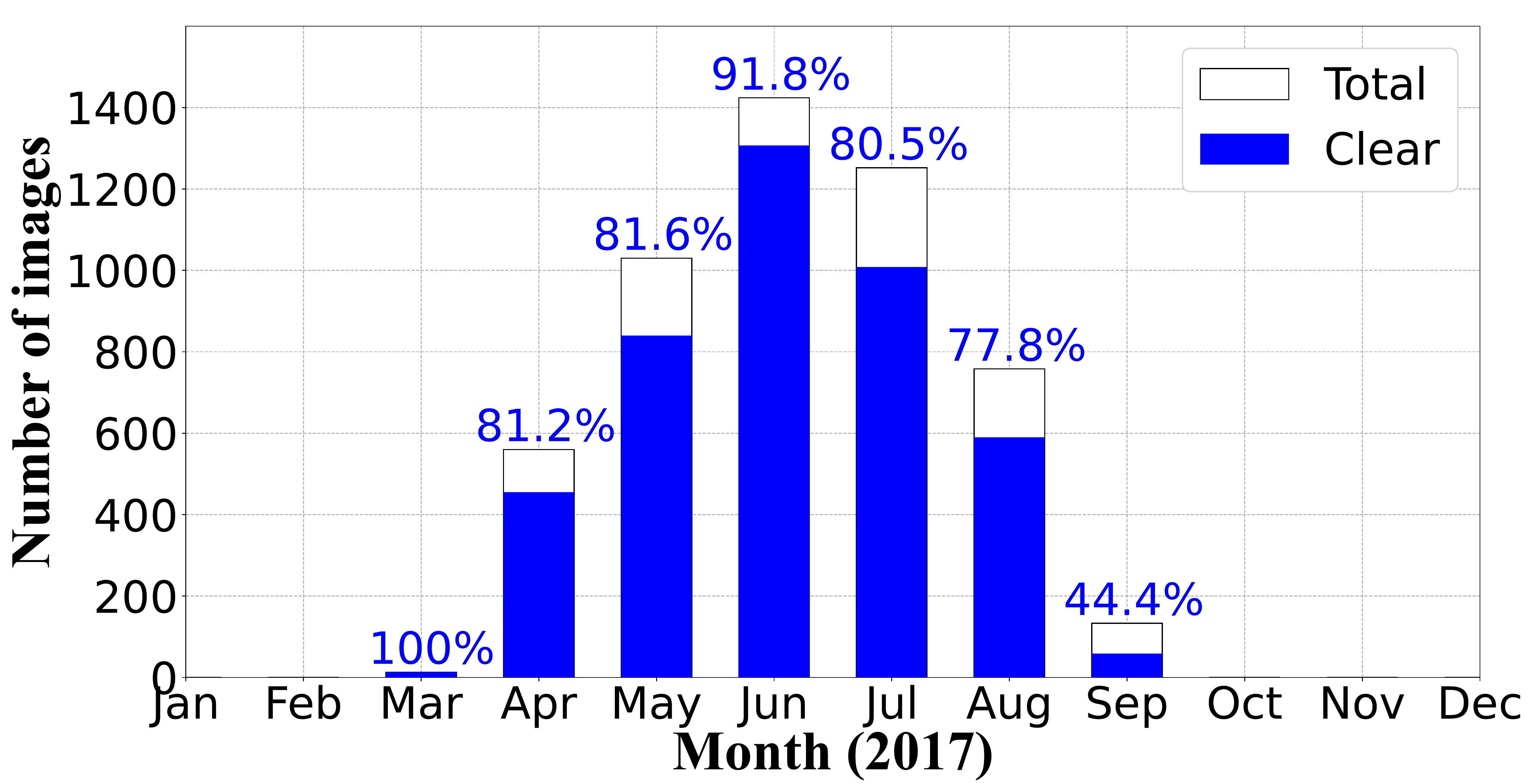}
 \caption{Monthly nighttime clear fraction in 2017. 
The number of total images is proportional to the relative length of
the nighttime in each month.
}
 \label{fig:month_compare}
\end{figure}

We compare the nighttime clear fraction in different months in 2017 in
Fig.~\ref{fig:month_compare}.  Only images that were taken when the sun was
13\degr\ below the horizon are shown.  As indicated in
Fig.~\ref{fig:datapoints}, we only had 24 hour non-stop observations
in June and early July.
It is good to see that as nighttime gets longer, the clear night
fraction also increases, to more than 90 per cent in June.  However,
the second half of the winter season does not seem to be as good as
the first half, even excluding September when the number of images
is small for statistics.

\begin{figure*}
 \includegraphics[width=\textwidth]{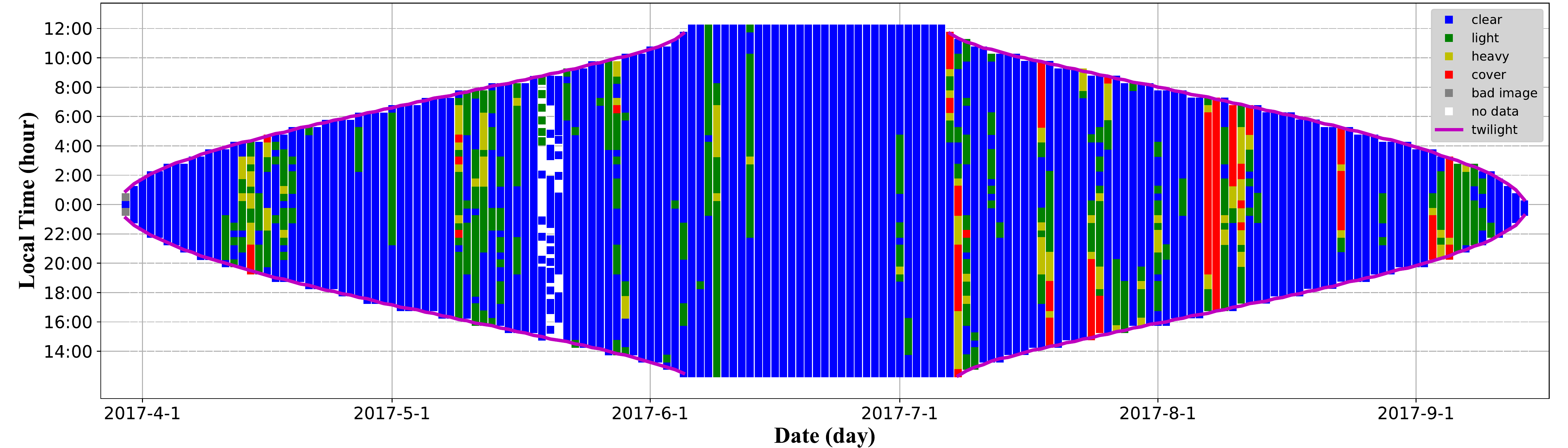}
 \caption{
The classification of every nighttime image in 2017.  Each image is
plotted as a small square centred at the local time when it was
taken.  The irregular missing data during 3 days in May 2017 
were due to a software glitch of the Canon camera.
}
 \label{fig:nightdata}
\end{figure*}

To further investigate the monthly variation, we plot the
classification of every image in Fig.~\ref{fig:nightdata}.
It is clear that there were more worse cases in the second half
of the season.  

\begin{figure}
 \includegraphics[width=\columnwidth]{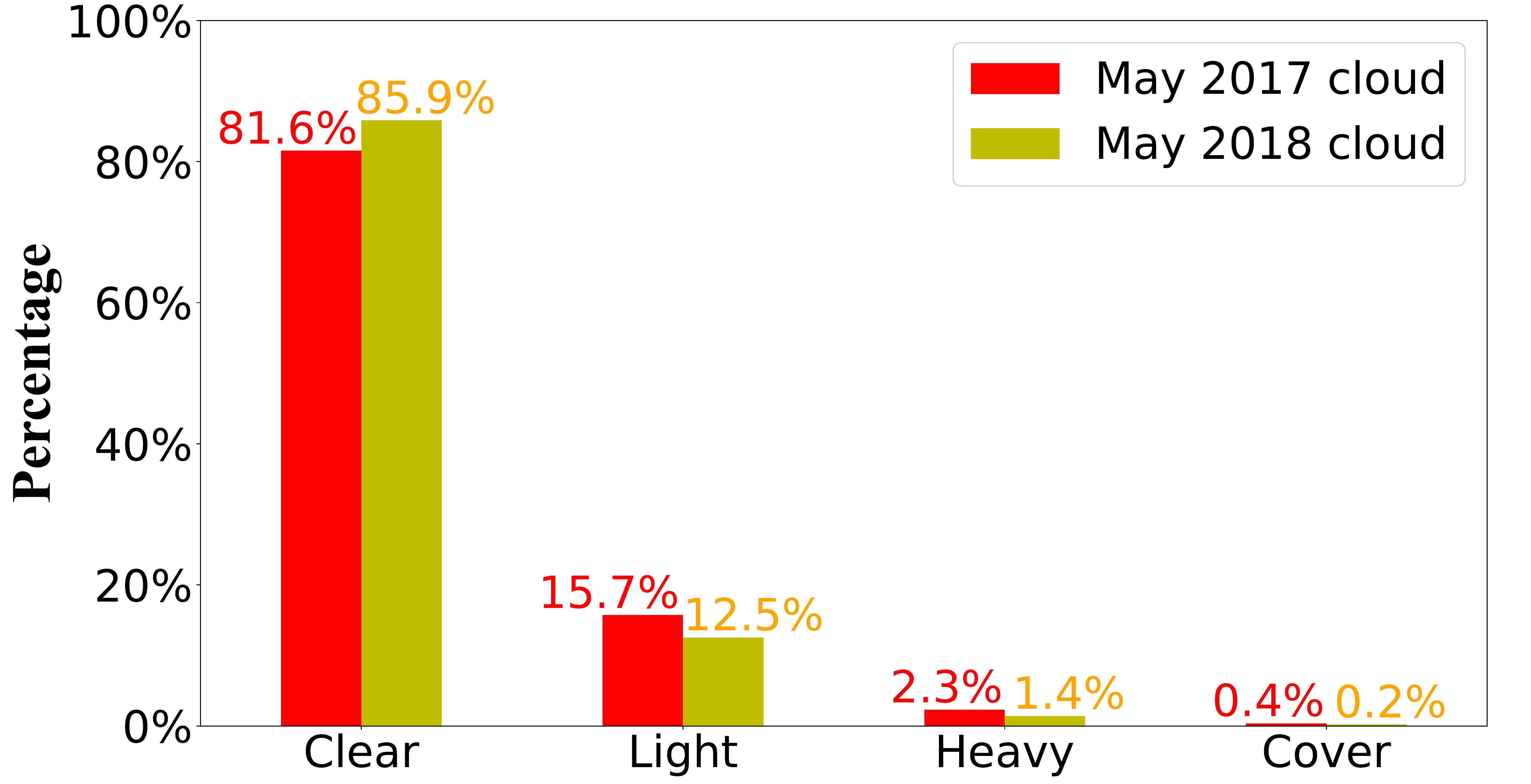}
 \caption{Comparison of the cloud statistics of May in 2017 and 2018.}
 \label{fig:may_compare}
\end{figure}

However, the above results are based only on data from 2017.
Since we have nighttime data in May for both 2017 and 2018, 
we made a simple comparison in Fig.~\ref{fig:may_compare}.
The statistics show similar but different results, with nearly 82 
and 86 per cent of `clear' class for 2017 and 2018, respectively;
and about 16 and 12 per cent of `light' class, respectively.
Therefore, we still cannot confirm or rule out any annual variation
and need long-term data to do so.

We are confident that our results are not caused by instrument effects, 
such as frosting on the lens. If frosting was an issue, the frost would 
build up gradually, but it is clearly not the case for the sudden cloud 
of `cover' class in July and August.  The design of KLCAM (Section~\ref{sec:ins}) 
has been proven to be very effective for preventing frosting.

Finally, we compare the nighttime clear fraction with those of TMT
candidate sites in Table~\ref{tab:compare}.  The 83 per cent clear
fraction at Dome~A seems to be the best among these sites.  
However, because of the high latitude, although Dome~A has
    the advantage of continuous dark time during polar nights, there is also the
    disadvantage of fewer total number of dark hours compared to
    those sites (Table~\ref{tab:compare}).  The annual total number
    of dark hours at Dome~A is only 49 per cent of that at Mauna
    Kea considering that the general astronomical night starts when the solar
    elevation angle is $-$18\,\degr, and it is 78 per cent when
    $-13$\,\degr\ is used for Dome~A \citep{zouhu}.

We also note that the result of this work is semi-quantitative, and we only have 1.5 years of data.

\begin{table}

 \caption{Comparison of clear fraction between Dome~A and the TMT 
 candidate sites \citep{asca}.
}
 \label{tab:compare}
 \centering
 \begin{tabular}{@{}lcccc@{}}
  \hline
  & & Dark & Clear & Clear\cr
  Site & Elevation & Hours$^{\rm a}$  & Fraction & Dark Hours$^{\rm b}$\\
  \hline
  Armazones & 3,064\,m & 3392\,hr & 82\%& 2798\,hr\\
  Tolar & 2,290\,m & 3390\,hr & 77\% & 2624\,hr\\
  Tolonchar & 4,475\,m & 3373\,hr & 72\% & 2442\,hr\\
  SPM & 2,830\,m & 3267\,hr & 73\% & 2372\,hr\\
  MK 13N & 4,050\,m & 3390\,hr & 71\% & 2404\,hr\\
  {\bf Dome~A} & {\bf 4,093\,m} & {\bf 2606\,hr} & {\bf83\%}& {\bf 2136\,hr} \\
  \hline
 \end{tabular}
 \flushleft
 $\rm ^a$ This refers to the number of annual dark hours defined according to the astronomical
 twilight  when the solar elevation angle is below $-18$\degr\ for TMT 
 sites and $-13$\degr\ for Dome~A (Section~\ref{sec:data}).\\
 $\rm ^b$ The annual clear dark hours are estimated from the clear fraction and the annual dark
 hours of each site.
 
\end{table}

\section{Discussion}
\label{sec:dis}

As presented in Section~\ref{sec:res}, although semi-quantitative, our
results show good agreement between our five independent analysers.  However, visual analysis is still
somewhat subjective.  In principle, one can even try to tell from a visual inspection of
the images whether a night was photometric or spectroscopic, but that
would be too subjective and we did not try it.
To partially eliminate subjective factors, machine learning and deep
learning methods can be employed \citep{Mommert}.  
However, a more complete approach would be to do direct photometry on
the all-sky images as mentioned in \citet{hrcam} and \citet{simsthesis}, measuring
various parameters for quantitative evaluation of the night sky, such
as sky background for aurora contamination, extinction and number of
stars for cloud cover.

Once established, these methods will be more efficient and can be
applied repeatedly as needed to a future larger dataset.  For 
the dataset of this work, we expect the new methods to generate more quantitative
results, but do not expect our general conclusions to change since the current visual
analysis produces relatively robust results as indicated by the
confusion matrices.

However, we notice that moonlight could add some uncertainty in our results.  
It could mostly affect the distinction between the `clear' and `light' classes, 
and has little effect on the `heavy' or `cover' classes in which most stars 
would disappear and there is no confusion.  Bright moonlight helps to spot `light' 
clouds, and this could possibly result in an underestimate of clouds for moonless nights 
when `light' clouds cannot be detected visually. On the other hand, it is also 
true that when the sky background is high with moonlight, fainter stars are hard 
to detect in visual inspection and thus a clear sky could be misclassified as 
`light' as most analysers are subjectively conservative and strict for `clear' class.  
This case is even worse when there were low clouds outside the 65\degr\ zenith angle 
circle while the moon was always low, never reaching 30\degr\ above the horizon during 
that period.  In this scenario, an overestimate of clouds could happen.  These 
complications are also reflected in the confusion matrices where the disagreement 
in the `light' classification is the greatest, about 16 per cent.  Moreover, in the 
result class of `light', there is a 10.7 per cent chance of misclassification towards 
`clear', incurring an uncertainty of 1.2 per cent for the `light' fraction of 
11.2 per cent (Fig.~\ref{fig:result}).  Similarly, in the result class of `clear', 
there is a 1.7 per cent chance of misclassification towards `light', giving an 
uncertainty of 1.4 per cent for the clear fraction of 83.3 per cent. Quantitative 
analyses mentioned above would eventually get rid of the subjective factors.

Our results show that aurorae could be detected nearly 45 per cent of the time, 
but they were strong for only 9 per cent of the time.  
One of the most common and strongest auroral emission lines is the 
[O\,{\sc i}]~557.7\,nm line, which is included in the standard $V$ band.
\citet{nigel} studied the aurorae at Dome~A with Nigel and Gattini data, and found that
the median auroral contribution to the $V$~band sky brightness is
 23.4~mag~arcsec$^{-2}$, while the median value of moonless sky brightness 
at night is 21.4~mag~arcsec$^{-2}$ \citep{gattini-cloud}.
This indicates roughly the extent to which auroral emission lines contribute to 
the broad band sky background.
However,  as discussed in Section~\ref{sec:aurora}, the strong auroral emission 
lines can be excluded by customized filters to minimize the contamination.
It has been reported that the aurora contribution to sky brightness at 
the South Pole can decrease by 1.6~$V$~mag statistically with a notch 
filter at 557.7~nm, demonstrating the efficiency of specially designed filters
\citep{dempsey}.

\begin{figure}
 \includegraphics[width=\columnwidth]{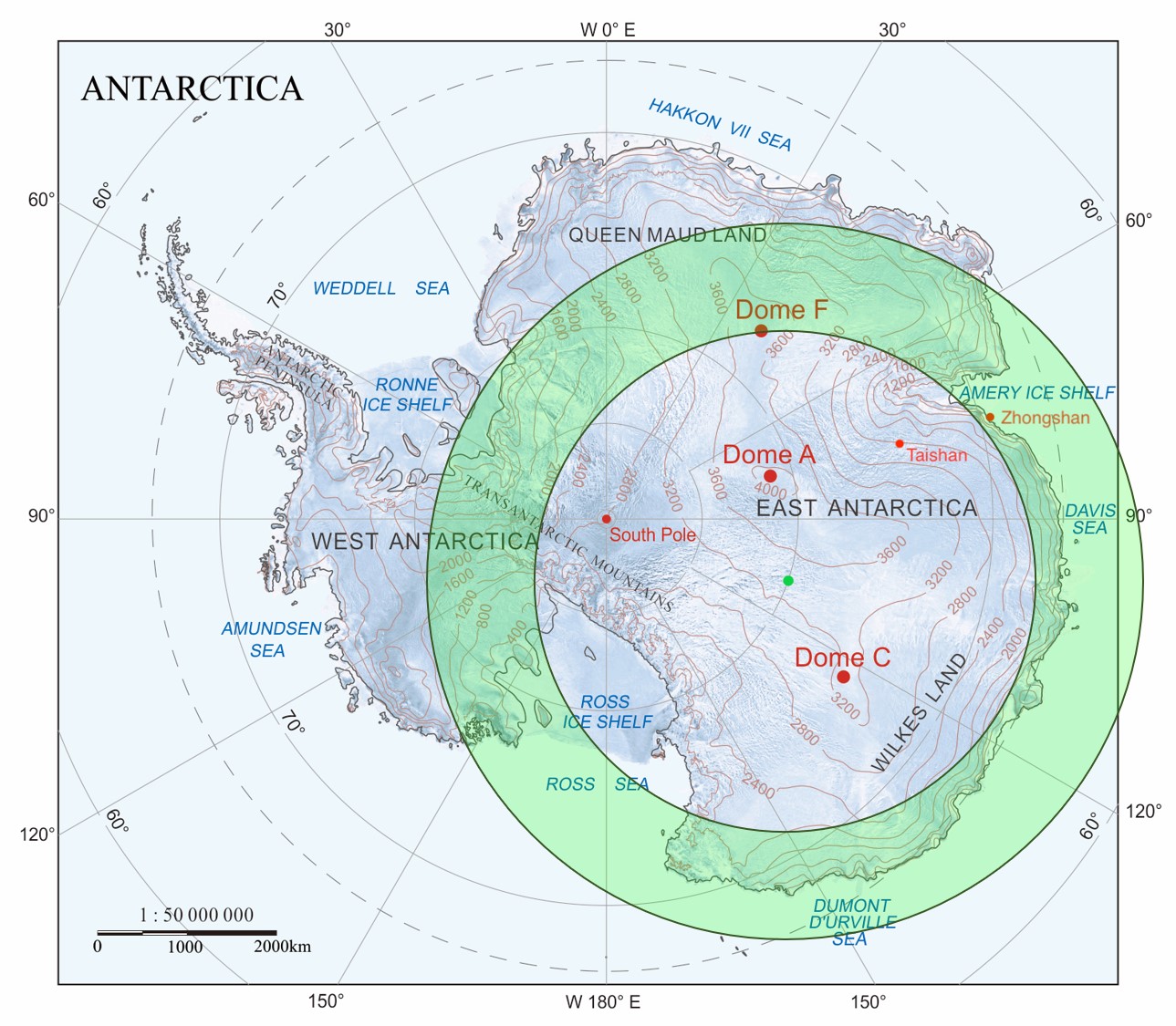}
 \caption{Illustration of the approximate location of the `auroral oval'
centred on the South Geomagnetic Pole (green dot) in Antarctica.  
Its location varies and does not have a fixed boundary.
Original map courtesy of Xiaoping Pang and Shiyun Wang.
}
 \label{fig:auroraloval}
\end{figure}

We also note that aurorae are stronger and appear more in one
direction (west to southwest) than the opposite direction.  This is because aurorae are
mostly from the `auroral oval' which is a ring shape area centred on
the South Geomagnetic Pole as illustrated in
Fig.~\ref{fig:auroraloval}.  Dome~A happens to be located inside the
auroral oval and so is usually not directly hit by aurorae from the zenith.  This
is a fortunate situation, and means that visible aurorae generally lie below or
close to the horizon at Dome~A, leaving less contamination in
the sky areas of low airmass than some other Antarctic sites.

Another issue is that when classifying the aurora in
Section~\ref{sec:res}, some images might only have little faint and thin
aurorae seen inside the `inner' area.  In such a case, we still mark
it as `inner', the second-worst class, because it would be purely subjective
to ignore it by judging its size, or mark it to a different class.
Moreover, since
we can see the faint aurora, it is possible that the whole area is all
contaminated by weaker aurorae, as is a reasonable inference considering
the sensitivity of the camera and human eyes.  However, in order to
keep objective as much as possible, we stuck to what can be seen by
human eyes, instead of even reasonable guessing.  
Future methods discussed above, such as photometry, can also solve this
problem.

Finally, KLCAM has an extra function that it can provide real-time
observing conditions over the whole sky, and therefore can help to
optimize observing plans of the telescopes at Dome~A, especially for
sky surveys.

\section{conclusion}

We present the statistical results of the cloud cover and aurora
contamination at Dome~A using 2017--2018 data from a fully functional
all-sky camera KLCAM. 

\begin{enumerate}

\item KLCAM was specifically designed and built for the harsh
environment at Dome~A, Antarctic.    It worked well unattended for 490
days since its installation in Jan. 2017 and took 47,035 images.

\item In total 6,664 nighttime images were visually inspected by 5
individual analysers and classified into 4 classes each for cloud cover
and aurora contamination.

\item We find 83 per cent of time with clear night sky at Dome~A,
and 55 per cent of time with no aurora at night.  Both of these
results are reliable with an uncertainty of about 2 per cent based on
the confusion matrices. 

\item The clear fraction at Dome~A is slightly better than that at the
best candidate site of TMT.

\item Long-term monitoring is still needed, and more objective 
analysing methods will be developed to obtain more quantitative 
systematic results.

\end{enumerate}

\section*{Acknowledgements}
\addcontentsline{toc}{section}{Acknowledgements}

We thank the CHINARE for their great efforts in installing
KLCAM, maintaining KLCAM and PLATO-A, and retrieving data.
This work has been supported by the National Natural
Science Foundation of China under Grant Nos. 11733007, 11673037,
11403057, and 11403048, the Chinese Polar Environment Comprehensive
Investigation and Assessment Programmes under grant No.
CHINARE2016-02-03, and the National Basic Research Program of China
(973 Program) under Grant No. 2013CB834900. PLATO-A is supported by the
Australian Antarctic Division.

\section*{Data Availability}
\addcontentsline{toc}{section}{Data Availability}

The data underlying this article will be shared on reasonable request to the corresponding author.







\bsp	
\label{lastpage}
\end{document}